\documentclass[aps,pre,twocolumn,superscriptaddress, nobibnotes,longbibliography]{revtex4-2}
\usepackage{amsmath,amssymb}
\usepackage{graphicx}
\usepackage{hyperref}
\usepackage{physics}
\usepackage{bm}
\usepackage{float}
\usepackage{caption}
\usepackage{subcaption}

\begin{document}

\title{X-ray photon correlation spectroscopy of hydrated lysozyme at elevated pressures}

\author{Milla Åhlfeldt}
\thanks{These authors have contributed equally to this work}
\affiliation{Department of Physics, AlbaNova University Center, Stockholm University, S-106 91 Stockholm, Sweden}

\author{Maddalena Bin}
\thanks{These authors have contributed equally to this work}
\affiliation{Department of Physics, AlbaNova University Center, Stockholm University, S-106 91 Stockholm, Sweden}

\author{Anita Girelli}
\affiliation{Department of Physics, AlbaNova University Center, Stockholm University, S-106 91 Stockholm, Sweden}

\author{Iason Andronis}
\affiliation{Department of Physics, AlbaNova University Center, Stockholm University, S-106 91 Stockholm, Sweden}

\author{Aigerim Karina}
\affiliation{Department of Physics, AlbaNova University Center, Stockholm University, S-106 91 Stockholm, Sweden}

\author{Nimmi Das Anthuparambil}
\affiliation{Deutsches Elektronen-Synchrotron, Notkestrasse 85, 22607 Hamburg, Germany}
\affiliation{Department Physik, Universität Siegen, Walter-Flex-Strasse 3, 57072 Siegen, Germany}

\author{Fiona Berner}
\affiliation{Institute for Physics, Johannes Gutenberg University Mainz, Staudingerweg 10, 55128 Mainz, Germany}
\affiliation{Max Planck Institute for Polymer Research, Ackermannweg 10, 55128 Mainz, Germany}

\author{Tobias Eklund}
\affiliation{Institute for Physics, Johannes Gutenberg University Mainz, Staudingerweg 10, 55128 Mainz, Germany}
\affiliation{Max Planck Institute for Polymer Research, Ackermannweg 10, 55128 Mainz, Germany}

\author{Louisa E. Kraft}
\affiliation{Institute for Physics, Johannes Gutenberg University Mainz, Staudingerweg 10, 55128 Mainz, Germany}
\affiliation{Max Planck Institute for Polymer Research, Ackermannweg 10, 55128 Mainz, Germany}

\author{Aliaksandr Leonau}
\affiliation{Department Physik, Universität Siegen, Walter-Flex-Strasse 3, 57072 Siegen, Germany}

\author{Fabian Westermeier}
\affiliation{Deutsches Elektronen-Synchrotron, Notkestrasse 85, 22607 Hamburg, Germany}

\author{Michael Sprung}
\affiliation{Deutsches Elektronen-Synchrotron, Notkestrasse 85, 22607 Hamburg, Germany}

\author{Christian Gutt}
\affiliation{Department Physik, Universität Siegen, Walter-Flex-Strasse 3, 57072 Siegen, Germany}

\author{Katrin Amann-Winkel}
\affiliation{Institute for Physics, Johannes Gutenberg University Mainz, Staudingerweg 10, 55128 Mainz, Germany}
\affiliation{Max Planck Institute for Polymer Research, Ackermannweg 10, 55128 Mainz, Germany}

\author{Fivos Perakis}
\email{f.perakis@fysik.su.se}
\affiliation{Department of Physics, AlbaNova University Center, Stockholm University, S-106 91 Stockholm, Sweden}

\begin{abstract}
Pressure provides a powerful parameter to control the protein conformation state, which at sufficiently high values can lead to unfolding. Here, we investigate the effects of increasing pressure up to $0.4$ GPa on hydrated lysozyme proteins, by measuring the nanoscale stress relaxation induced and probed by X-rays. Structural and dynamical information at elevated pressures was obtained using X-ray photon correlation spectroscopy (XPCS) in combination with a diamond anvil cell (DAC). The dynamical analysis revealed a slowing down of the system up to $0.2$ GPa, followed by a re-acceleration at $0.4$ GPa. A similar non-monotonic behavior was observed both in the Porod and Kohlrausch-Williams-Watts (KWW) exponents, consistently indicating a crossover between $0.2$ and $0.4$ GPa. These findings suggest the presence of pressure-induced structural changes that impact protein collective stress-relaxation as the system transitions from a jammed state to an elastically driven regime. These results may be relevant for a deeper understanding of protein stability under compression as well as for practical high-pressure technologies, including food processing and pharmaceutical applications.
\end{abstract}

\maketitle

\section{Introduction}

The ability to manipulate and process proteins has become increasingly important due to their expanding use in medicine, such as in vaccines, antibodies and enzymes \cite{ebrahimi_engineering_2023}, as well as in food preservation \cite{miron_study_2021}. Pressure is a fundamental thermodynamic variable with a wide range of applications involving proteins \cite{akasaka_proteins_2022, moron_gelation_2022}, in particular high-pressure food processing for which pressures up to approximately $0.6$ GPa are routinely employed \cite{balasubramaniam_Highpressure_2008}. High-pressure processing provides a means to extend the shelf life of food products without the use of heat or additives and preservatives \cite{balasubramaniam_Highpressure_2008, nabi_Highpressure_2021}. Similarly to temperature, sufficiently high pressures can induce denaturation of proteins \cite{silva_pressure_2001}. Studies indicate that the primary mechanism behind pressure-induced unfolding is the weakening of the hydrophobic effect and the reduction of internal cavities in the folded protein state, facilitated by the penetration of water molecules into hydrophobic regions, ultimately leading to destabilization \cite{hummer_pressure_1998, grigera_behavior_2010, roche_cavities_2012}.

The pressure-temperature phase diagram of proteins is characterized by an ellipsoidal stability region within which the protein retains its native conformation, whereas denaturation occurs outside this region \cite{messens_use_1997, smeller_Pressure_2002}. In addition to the native and unfolded states, intermediate molten globule states can also exist. These states exhibit structural features that are partially retained from the native conformation but lack biological function \cite{gupta_Premolten_2023, acharya_Dry_2022, masson_Pressureinduced_1996}.

The structure and function of biomolecules are strongly influenced by the layers of surrounding water molecules, often referred to as hydration water, which can interact directly with the protein surface \cite{laage_water_2017, roh_influence_2006}. As protein stability and dynamics are closely coupled to the properties of the hydration layer, studying hydrated proteins is essential for understanding both protein behavior and the role of hydration water in biomolecular processes \cite{swenson_properties_2007, schiro_communication:_2013}. Moreover, hydrated protein samples provide a unique opportunity to probe the dynamics of hydration water itself \cite{bagchi_water_2005, bin_Wideangle_2021}.

To investigate the pressure-dependence of nanoscale stress-relaxation of hydrated proteins, X-ray photon correlation spectroscopy (XPCS) provides a powerful experimental technique capable of probing dynamics across timescales ranging from minutes to nanoseconds \cite{perakis_molecular_2020, madsen_simple_2010,grubel_Xray_2008}. High-pressure XPCS has emerged as a unique capability that benefits from the high coherence of fourth-generation synchrotron sources~\cite{Cornet:vl5019}. XPCS quantifies fluctuations in scattering intensities caused by changes in the spatial configuration of particles within the illuminated sample volume, including diffusive motions and structural rearrangements. The corresponding relaxation times of these fluctuations reflect the characteristic timescales of particle dynamics. Information on inter-particle correlations and their time evolution, related to the fluctuations, are contained in the intermediate scattering function $F(q,t)$. The measurement yields the intensity autocorrelation function $g_2$, by correlating over time $t$, for different momentum transfer vectors $q$, which is related to the intermediate scattering function through the Siegert relation \cite{voigt_comparison_1994}. 

Recent work has demonstrated the utility of XPCS for studying nanoscale stress-relaxation in hydrated lysozyme over a broad temperature range from ambient to cryogenic conditions \cite{bin_coherent_2023}. By varying the incident beam flux, the observed relaxation times could be associated with nanoscale stress-relaxation stimulated by the X-ray beam, which reflects the intrinsic dynamic viscoelastic response of the system to external stimulus. This approach resembles conceptually other experimental methods that probe the changes under some external force, such as cyclic shearing \cite{dauchot_dynamical_2005}, used in granular media close to the “jamming  transition”, as well as dielectric spectroscopy used to probe the dynamic response driven by external fields \cite{albert_Fifthorder_2016}. Furthermore, complementary small- and wide-angle X-ray scattering measurements indicated no major structural changes, suggesting that beam-induced dynamics do not necessarily lead to structural degradation and loss of function \cite{bin_coherent_2023}.

Here, we investigate the X-ray induced dynamical response of hydrated lysozyme \cite{bin_coherent_2023} at elevated pressures, ranging from ambient pressure up to $0.4$ GPa using XPCS in ultra small-angle X-ray scattering (USAXS) geometry. A diamond anvil cell (DAC) was employed to achieve elevated pressures, enabling the direct examination of pressure effects. By combining XPCS with the DAC, we were able to probe the structural and dynamical effects in the hydrated protein clusters upon compression.

\section{Methods}

\subsection{Sample Preparation}
The lysozyme powder used in this study was derived from hen egg white and manufactured by Roche (product number 10837059001). Prior to hydration, the powder was ground using a mortar to reduce the grain size, but no additional purification or drying steps were performed. Hydration was achieved using a controlled hydration chamber until the desired hydration level, $h=\frac{m_{water}}{m_{protein}} \approx 0.3$, was attained. This hydration level ensures the protein molecules are fully covered by water molecules and able to maintain biological activity \cite{bellissent-funel_water_2016}. During hydration, the dry powder was placed inside the closed hydration chamber which allowed regulation of both the relative humidity and temperature. The hydration level was determined gravimetrically by weighing the protein powder before and after the hydration process.

\subsection{X-ray Photon Correlation Spectroscopy}
The experiments were conducted at the P10 beamline at PETRA III at the Deutsches Elektronen-Synchrotron (DESY). All measurements were performed in ultra-small angle X-ray scattering (USAXS) geometry, with a sample-detector distance (SDD) of $21.2$ m and at a photon energy of $12$ keV. Scattering patterns were recorded using an Eiger X 4M detector with an exposure time of $0.5$ s per frame. For most measurements, $2000$ frames were collected. All measurements were performed at approximately $295$ K. A summary of the experimental parameters is provided in Table \ref{tab:exp_param}.

Dynamic information was extracted from the XPCS measurement by analyzing temporal correlations of intensity fluctuations in the scattering patterns \cite{perakis_molecular_2020}. For a given momentum transfer \textit{q}, the normalized intensity autocorrelation function, $g_2(q,t)$, is defined as
\begin{equation}
g_2(q,t) = \frac{\langle I(q,t_0) I(q,t_0+t) \rangle}{\langle I(q,t_0) \rangle^2},
\end{equation}
where $I(q,t_0)$ is the intensity at time $t_0$ and $\langle \cdots \rangle$ denotes averaging over $t_0$. The correlation functions $g_2(q,t)$ can be fitted with a stretched exponential function:
\begin{equation} \label{eq:stretched_exp}
g_2(q,t)= \beta(q) e^{-2[\Gamma(q) t]^{\alpha(q)}} + B,
\end{equation}
where $\beta$ is the speckle contrast, $\Gamma$ is the relaxation rate, $\alpha$ is the Kohlrausch-Williams-Watts (KWW) exponent and $B$ is the baseline. The KWW exponent provides information on the nature of the dynamics in the system. An exponent $\alpha=1$ indicates diffusive Brownian motion, while $\alpha<1$ usually suggests dynamical heterogeneity and $\alpha > 1$ is associated with stress-driven dynamics \cite{ishii_spatial_2021,frey_Liquidlike_2025}, and has been observed in disordered soft solids such as gels and creams \cite{fluerasu_slow_2007, herzig_dynamics_2009} as well as in glass forming liquids \cite{guo_nanoparticle_2009, caronna_dynamics_2008}.

To further probe the temporal evolution of the system, the two-time correlation (TTC) function $c_2(q,t_1,t_2)$, was computed. It is defined as
\begin{equation}
c_2(q,t_1,t_2) = \frac{\langle I(q,t_1)I(q,t_2)\rangle_{pix}}{\langle I(q,t_1)\rangle_{pix} \langle I(q,t_2)\rangle_{pix}},
\end{equation}
where $I(q,t)$ is the scattered intensity at momentum transfer \textit{q} and time $t$, and $\langle \cdots \rangle$ denotes averaging over all pixels corresponding to the selected \textit{q}-value. The two-time correlation function $c_2(q, t_1,t_2)$ allows to probe the evolutions of dynamics and is sensitive to underlying fluctuations. Therefore, it is particularly useful for determining time-dependent behavior such as aging effects and temporally heterogeneous dynamics in the system.

\begin{table}[H]
    \centering
    \caption{Experimental parameters}
    \begin{tabular}{|c|c|}
        \hline
       Photon energy  & 12 keV \\
       \hline
        Beam size & 35 x 35 $\mu$m$^2$\\
        \hline
        Flux & $2\cdot 10^9$ ph/s\\
        \hline
        Detector & Eiger X 4M\\
        \hline
        SDD & 21.2 m\\
        \hline
        Exposure time & 0.5 s\\
        \hline
    \end{tabular}
    \label{tab:exp_param}
\end{table}

\subsection{Diamond Anvil Cell}
The DAC consists of two opposing diamonds, each with a thickness of $1$ mm. The sample was placed inside a steel gasket positioned between the diamonds. The sample chamber has a window of $400$ $\mu$m in diameter. The applied pressure is typically determined via ruby fluorescence spectroscopy, as the wavelength of the ruby fluorescence line shifts in a well-characterized manner with pressure \cite{shen_Highpressure_2016}. In the present study, the reported pressures were calibrated based on previous work using the same DAC experimental setup with an estimated uncertainty of $\pm 0.02$ GPa \cite{karina_multicomponent_2025}. The total pressure uncertainty is estimated to be approximately $\pm 0.04$, accounting for both the DAC calibration uncertainty and the additional uncertainty introduced by manually applying pressure by tightening the screws to the cell. The measurement at 0.1 GPa possibly carries a larger uncertainty since it is at the limit of the DAC capabilities, as reflected in the corresponding individual $g_2$ functions (see Supporting Information). 

\section{Results and Discussion}

\subsection{Ultra Small-angle X-ray Scattering}
\begin{figure}[tbh]
    \centering
    \includegraphics[width=0.9\linewidth]{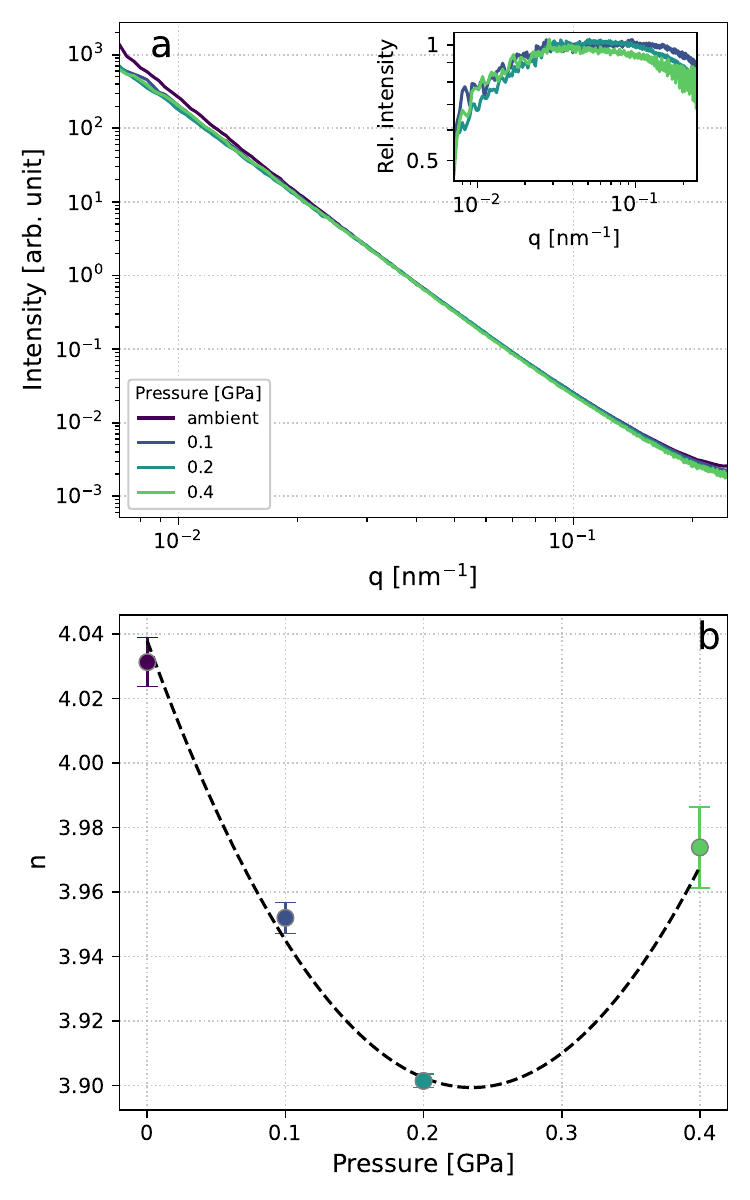}
    \caption{(a) Scattering intensities $I(q)$ as a function of momentum transfer $q$, averaged over several datasets for each pressure and normalized to the intensity at $q \approx 0.04$ nm$^{-1}$. The inset shows the relative intensities of all elevated pressures normalized to the intensity at ambient pressure. (b) The Porod exponents $n$, averaged over all individual exponents extracted using Eq. \ref{eq:porod} for each pressure. The error bars were calculated using the standard error. The dashed curve denote a parabolic fit to highlight the trend.}
    \label{fig:usaxs_all_mean}
\end{figure}

The scattering intensities as a function of momentum transfer are shown in Fig. \ref{fig:usaxs_all_mean}a for all pressures, averaged over several datasets. The intensity curves were fitted using Porod's law \cite{hammouda_new_2010}, 
\begin{equation} \label{eq:porod}
I(q)=Aq^{-n} + B, 
\end{equation}
where $A$ is the amplitude, $B$ is a baseline and $n$ the Porod exponent, which reveals information on surface roughness. The Porod fits for all individual measurements are shown in the Supporting Information. The Porod exponents at the different pressures are presented in Fig. \ref{fig:usaxs_all_mean}b.

The scattering intensity in the Porod range, which here is observed at approximately $q<0.15$ nm$^{-1}$, is attributed to scattering from grain boundaries or the solid protein-air interface \cite{phan-xuan_Hydrationinduced_2020, bin_coherent_2023}. In the present study, the Porod exponent $n$ decreases upon compression up to a pressure of $0.2$ GPa, followed by an increase at $0.4$ GPa. A Porod exponent of $n=4$ corresponds to smooth surfaces or well-defined interfaces, whereas $n=3$ indicates increased surface roughness \cite{hammouda_new_2010}. Therefore, the observed non-monotonic pressure dependence suggests that above a threshold pressure between $0.2$ and $0.4$ GPa, the protein network undergoes a transition that impacts the interfacial characteristics and modifies the local packing density.

\subsection{Intensity Autocorrelation Functions}

\begin{figure}[tbh]
    \includegraphics[width=1\linewidth]{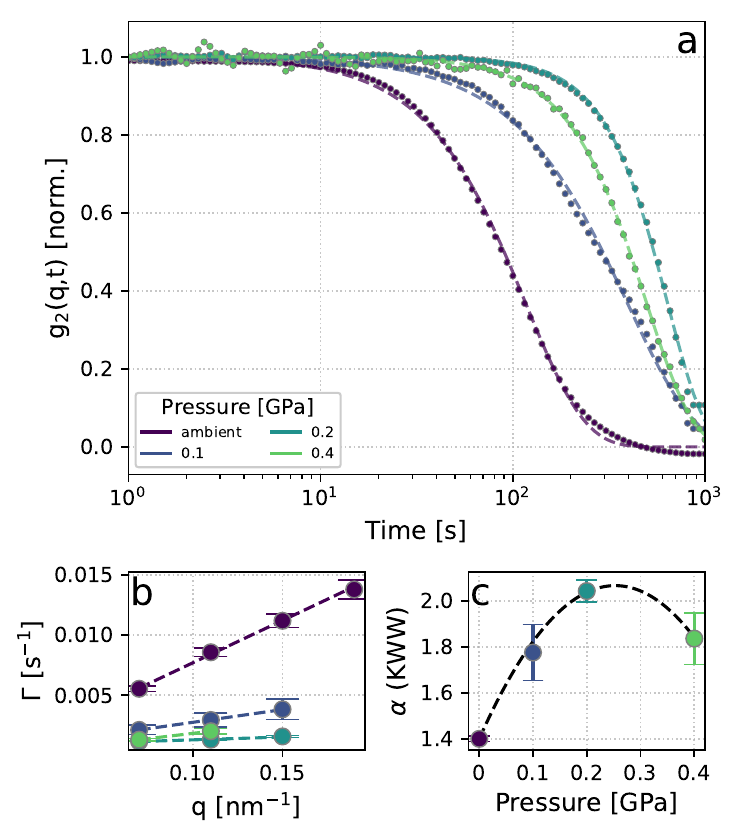}
    \caption{(a) The normalized intensity autocorrelation function $g_2(q,t)$ at momentum transfer $q=0.07$ nm$^{-1}$, with the baseline $B$ subtracted and normalized to the contrast $\beta$. All pressures are shown and have been averaged over several measurements at the same pressure. (b) Relaxation rates $\Gamma$, extracted from the $g_2$ functions, as a function of momentum transfer $q$, averaged over all measurements for each pressure and $q$-value. The dashed lines denote a linear fit made to the relaxation rates. (c) The Kohlrausch-Williams-Watts (KWW) exponents $\alpha$, extracted from the $g_2$ functions, as a function of pressure, averaged over all measurements for each pressure. Error bars are given by the standard error of the extracted values. The dashed curve denote a parabolic fit to highlight the trend.}
    \label{fig:xpcs_all_average}
\end{figure}

The intensity autocorrelation functions $g_2(q,t)$ for momentum transfer value $q = 0.07$ nm$^{-1}$ are shown in Fig. \ref{fig:xpcs_all_average}a for all measured pressures, averaged over several datasets (the individual measurements are provided in the Supporting Information). Previous studies on X-ray-induced stress-relaxation dynamics in hydrated proteins have demonstrated that this approach can probe systems whose equilibrium dynamics are otherwise inaccessible in the measurement time window, such as granular matter and glassy materials \cite{bin_coherent_2023}. In such systems, the relaxation has been attributed to the transition from a jammed granular state to an elastically driven regime. Similar X-ray-induced dynamics have also been reported in oxide glasses, where the X-ray beam was found to fluidize the glass through local reconfigurations that trigger larger-scale rearrangements due to internal stresses \cite{pintori_relaxation_2019}.

The dynamics of the system exhibit a significant pressure dependence. As seen in Fig. \ref{fig:xpcs_all_average}a, the $g_2$ slows progressively upon compression up to $0.2$ GPa, after which the system reaccelerates at $0.4$ GPa. Similar non-monotonic behavior has been reported in metallic glasses studied using X-ray photon correlation spectroscopy in combination with wide-angle X-ray diffraction under varying pressures \cite{zhang_Pressureinduced_2023}. In the metallic glass system, the relaxation dynamics initially slowed down upon compression, as expected, but beyond a threshold pressure the dynamics accelerated rapidly. In the same study, complementary wide-angle X-ray diffraction measurements revealed a pronounced local structural change at the same pressure, consistent with a pressure-induced polyamorphic transition from a low-density amorphous to a high-density amorphous state, relating the structural transition with the observed dynamical anomaly~\cite{zhang_Pressureinduced_2023}.

The corresponding relaxation rates $\Gamma$, extracted from the intensity autocorrelation functions and averaged over all measurements, are presented in Fig. \ref{fig:xpcs_all_average}b as a function of momentum transfer. The extracted relaxation rates were fitted with a linear function, denoted by the dashed lines in Fig. \ref{fig:xpcs_all_average}b. The observed linear $q$-dependence of $\Gamma$ is consistent with previous studies, which identified ballistic motion in hydrated lysozyme \cite{bin_coherent_2023}. Furthermore, the KWW exponent, shown in Fig. \ref{fig:xpcs_all_average}c, increases upon increasing pressure to $\alpha \approx 2$ at $0.2$ GPa, followed by a decrease at $0.4$ GPa. At ambient pressure, $\alpha \approx 1.4$, is in agreement with previous findings of $\alpha \approx 1.5$ in hydrated lysozyme \cite{bin_coherent_2023}, which is characteristic of driven dynamics.

Taken together, both the structural analysis, reflected in the Porod exponent $n$ from the scattering intensities, and the dynamical analysis, through the trend of the relaxation rates $\Gamma$ and KWW exponents $\alpha$, reveal a consistent non-monotonic pressure dependence. All observables indicate a crossover between $0.2$ and $0.4$ GPa.

\subsection{Two-time Correlation Functions}
\begin{figure}[H]
    \includegraphics[width=1.\linewidth]{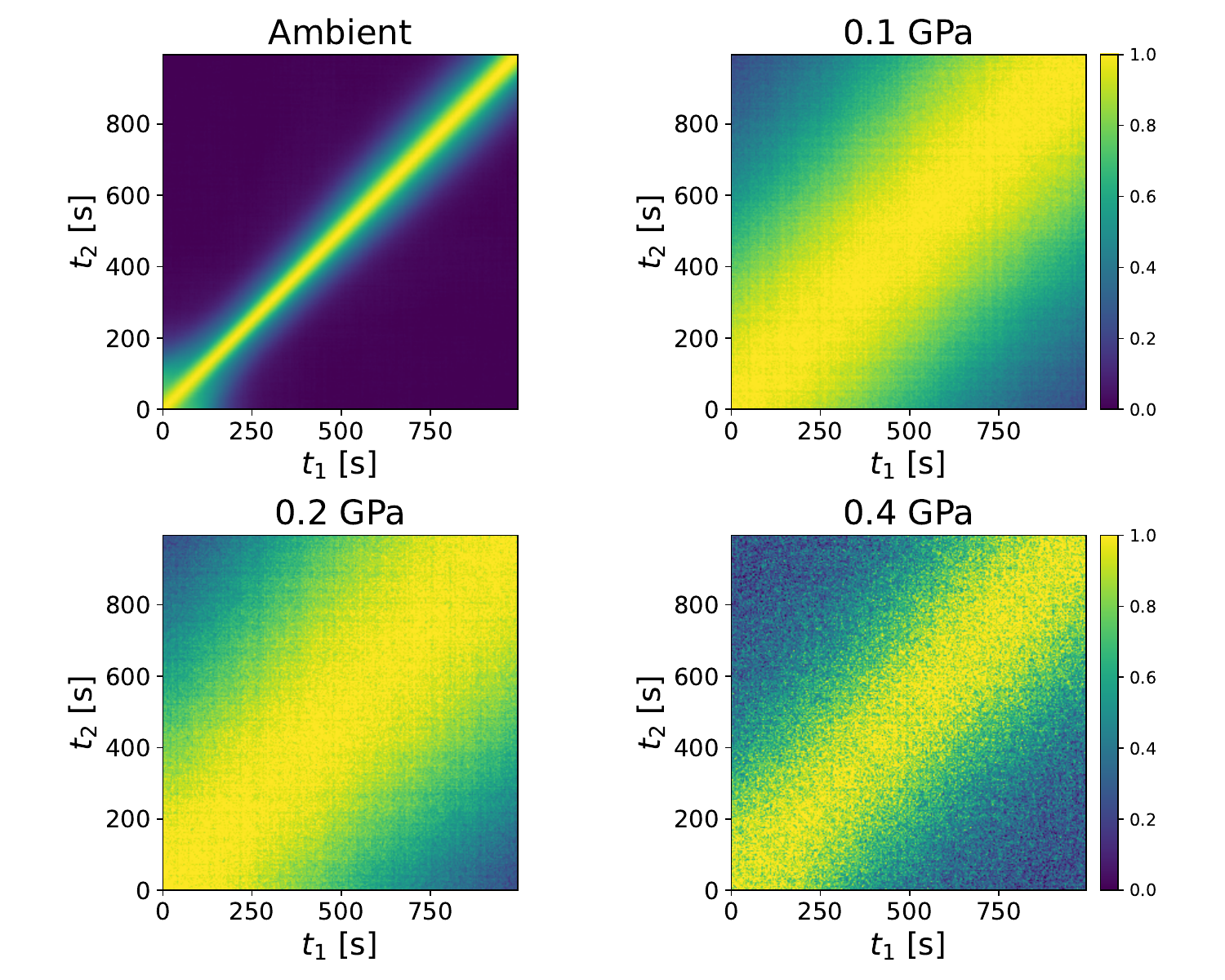}
    \caption{The two-time correlation functions with the baseline subtracted and normalized to the diagonal at momentum transfer $q=0.07$ nm$^{-1}$ for all pressures, as indicated by each panel header.}
    \label{fig:ttc}
\end{figure}

The two-time correlation (TTC) functions with the baseline subtracted and normalized to the diagonal for all measured pressures at momentum transfer $q=0.07$ nm$^{-1}$ are presented in Fig. \ref{fig:ttc}. One representative TTC function is displayed for each pressure to highlight the general trends, although some variation between individual measurements within each pressure condition remains, as also observed in the $g_2$ functions (see Supporting Information). The TTC functions reveal a relative slowdown of dynamics up to $0.2$ GPa. At $0.4$ GPa, the system appears to accelerate again, as indicated by the narrower diagonal compared to that observed at $0.2$ GPa. This trend is consistent with the results from the $g_2$ analysis and further supports the existence of a dynamical crossover in the range between $0.2$ and $0.4$ GPa. Moreover, the approximately constant width of the diagonal with respect to time at each pressure suggests stationary dynamics during the course of the measurements.

As previously discussed, similar behavior has been reported in metallic glasses \cite{zhang_Pressureinduced_2023}, where structural and dynamical changes were linked through a pressure-induced polyamorphic transition. Here, we observe a crossover in both the structural analysis, through the Porod exponents, and in the dynamical analysis, through the relaxation rates and KWW exponents. This could suggest that the hydrated lysozyme system undergoes a structural transition upon compression, accompanied by a shift into a faster dynamical regime. Unlike the metallic glass study, however, here we probe the nanoscale lengthscales (in the USAXS regime) related to the stress relaxation of the protein matrix. The observed crossover may therefore be related to a rearrangement in the local protein packing, as indicated by the Porod exponent pointing towards smoother surfaces and more well-defined interfacial characteristics.

While lysozyme in aqueous solution is expected to denature at pressures of several hundred MPa, irreversibly above $0.55$~GPa \cite{heremans_Pressure_1985}, studies suggest that the crystalline form may tolerate pressures up to $0.95$~GPa, likely due to crystal packing mitigating the effects of compression \cite{jonas_Highpressure_1990, yamada_Highpressure_2015}. Studies on the partial molar volume (PMV) of a protein, defined as the sum of the constituent atomic volumes, internal cavities and the change in volume because of hydration, suggested that upon compression, the system first shrinks due to compression of cavities, followed by a plateau in the PMV as the cavities are filled with water molecules from the bulk \cite{yamada_Highpressure_2015}.

Additional insight may be drawn from the hydration level-temperature phase diagram of lysozyme-water systems obtained from sorption and desorption calorimetry \cite{kocherbitov_lysozymewater_2004}. The phase diagram demonstrates transitions from rigid to more flexible states and indicates that upon heating, the onset of increased flexibility occurs at lower hydration levels. Therefore, the present results suggest that pressure, like temperature, may act as a parameter that shifts hydrated lysozyme from a rigid state to a more flexible regime, mediated by changes in hydration and structural packing.

\section{Conclusions}
Summarizing, the structural and dynamical response of hydrated lysozyme upon compression were measured using X-ray photon correlation spectroscopy in ultra small-angle X-ray scattering geometry, in combination with a diamond anvil cell that enabled in situ measurements up to $0.4$ GPa. Structural information was obtained from the scattering intensities as a function of momentum transfer, from which Porod exponents were extracted. Dynamical behavior was characterized through intensity autocorrelation functions, yielding relaxation rates and KWW exponents, while two-time correlation functions provided additional temporal resolution.

This method allowed us to probe the dynamics of hydrated lysozyme using X-rays, consistent with previous work on the same system at ambient pressure \cite{bin_coherent_2023}. These dynamics have been associated with a transition from a jammed granular state to an elastically driven regime, demonstrating that XPCS can access dynamical information in systems with otherwise inaccessible equilibrium dynamics.

Both structural and dynamical analyses revealed a non-monotonic pressure dependence, with a consistent crossover observed between $0.2$ and $0.4$ GPa. Up to $0.2$ GPa, the dynamics slow down, consistent with pressure-induced compaction, while at $0.4$ GPa the system re-accelerated. A similar anomaly has been observed in metallic glasses under pressure \cite{zhang_Pressureinduced_2023}, where the crossover was linked to a polyamorphic transition from a low-density amorphous to a high-density amorphous state. The observed crossover here in both the structural and dynamical analyses may indicate a correlation between rearrangements in the local protein packing and the stress relaxation. Furthermore, comparison with hydration level-temperature phase diagrams of lysozyme-water systems \cite{kocherbitov_lysozymewater_2004}, may suggest that pressure, analogous to temperature, can drive a transition from a rigid to a more flexible state, induced by rearrangements in the local packing.

In conclusion, these results demonstrate that XPCS in combination with high-pressure techniques provides a powerful tool for probing hydrated protein dynamics under pressure. Such insights into protein behavior upon compression may not only advance the fundamental understanding of protein-water interactions, but may also be useful for advances in high-pressure food processing as well as high-pressure medical applications.

\section*{Acknowledgments}
We would like to acknowledge financial support by the Swedish National Research Council (Vetenskapsrådet) under Grant No. 2019-05542, 2023-05339, by the Knut och Alice Wallenberg foundation (WAF, Grant. No. 2023.0052) and from the European Union’s Horizon Europe research and innovation programme under the Marie Skłodowska-Curie grant agreement No. 101081419 (PRISMAS) (F.P. \& I.A.) and 101149230 (CRYSTAL-X) (F.P. \& A.G.). We acknowledge Deutsches Elektronen-Synchrotron DESY (Hamburg, Germany), a member of the Helmholtz Association HGF, for the provision of experimental facilities. Beamtime was allocated for proposal I-20230752 EC at beamline P10 at PETRA III. We thank Iris Schwark from the PETRA III sample environment group, as well as Hans-Peter Liermann (DESY) for discussion and technical help with our DAC setup. L.E.K., T.E., K.A.-W acknowledge funding by the Deutsche Forschungsgemeinschaft (DFG, German Research Foundation) in the framework to the collaborative research center "Defects and Defect Engineering in Soft Matter" (SFB1552) under Project No. 465145163 (subproject Q3). C.G. acknowledges funding from BMBF (05K19PS1, 05K20PSA and 05K22PS1). C.G. acknowledges funding from NFDI 40/1 (DAPHNE4NFDI). This
work was supported through the Maxwell computational resources
operated at Deutsches Elektronen-Synchrotron DESY, Hamburg, Germany. We thank the Centre for Molecular Water Science (CMWS) for scientific exchange and support.

\bibliography{Hydlyspress.bib}

\end{document}